\documentclass[apjl]{emulateapj}

\newcommand{\elecd}{$n_{\rm e}$}
\newcommand{\te}{$T_{\rm e}$}

\newcommand{\ha}{H$\alpha$}
\newcommand{\fnii}{[N~{\sc ii}]}

\newcommand{\foiii}{[O~{\sc iii}]}
\newcommand{\fneiii}{[Ne~{\sc iii}]}
\newcommand{\niii}{N~{\sc iii}}
\newcommand{\ciii}{C~{\sc iii}}
\newcommand{\oii}{O~{\sc ii}}
\newcommand{\cii}{C~{\sc ii}}
\newcommand{\neii}{Ne~{\sc ii}}
\newcommand{\hi}{H\,{\sc i}}
\newcommand{\hii}{H~{\sc ii}}

\newcommand{\adf}{{\it adf}}
\newcommand{\adfs}{{\it adf}s}

\shorttitle{Imaging of NGC\,6778}
\shortauthors{Garc\'{\i}a-Rojas et al.}

\begin{document}


\title{Imaging the elusive H-poor gas in the high {\adf} planetary nebula NGC\,6778}


\author{Jorge Garc\'{\i}a-Rojas\altaffilmark{1,2} and Romano L. M. Corradi\altaffilmark{3,1} and Hektor Monteiro\altaffilmark{4} and David Jones\altaffilmark{1,2} and Pablo Rodr\'{\i}guez-Gil\altaffilmark{1,2} and Antonio Cabrera-Lavers\altaffilmark{3,1}}





\altaffiltext{1}{Instituto de Astrof\'{\i}sica de Canarias, E-38200, La Laguna, Tenerife, Spain}
\altaffiltext{2}{Universidad de La Laguna. Depart. de Astrof\'{\i}sica, E-38206, La Laguna, Tenerife, Spain}
\altaffiltext{3}{GRANTECAN, Cuesta de San Jos\'e s/n, E-38712 , Bre\~na Baja, La Palma, Spain}
\altaffiltext{4}{Instituto de F\'{\i}sica e Qu\'{\i}mica, Universidade Federal de Itajub\'a, Av. BPS 1303-Pinheirinho, 37500-903, Itajub\'a, Brazil}

\begin{abstract}
We present the first direct image of the high-metallicity gas component in a planetary nebula (NGC\,6778), taken with the OSIRIS Blue Tunable Filter centered on the {\oii} $\lambda$4649+50 \AA\  optical recombination lines (ORLs) at the 10.4m Gran Telescopio Canarias. We show that the emission of these faint {\oii} ORLs is concentrated in the central parts of the planetary nebula and is not spatially coincident either with emission coming from the bright {\foiii} $\lambda$5007 \AA\  collisionally excited line (CEL) or the bright {\ha} recombination line. From monochromatic emission line maps taken with VIMOS at the 8.2m Very Large Telescope, we find that the spatial distribution of the emission from the auroral {\foiii} $\lambda$4363 line resembles that of the {\oii} ORLs but differs from nebular {\foiii} $\lambda$5007 CEL distribution, implying a temperature gradient inside the planetary nebula. The centrally peaked distribution of the {\oii} emission and the differences with the {\foiii} and {\hi} emission profiles are consistent with the presence of an H-poor gas whose origin may be linked to the binarity of the central star. However, determination of the spatial distribution of the ORLs and CELs in other PNe, and a comparison of their dynamics is needed to further constrain the geometry and ejection mechanism of the metal-rich (H-poor) component and hence, understand the origin of the abundance discrepancy problem in PNe.
\end{abstract}


\keywords{planetary nebulae: individual
--- ISM: abundances --- binaries: close}



\section{Introduction}

In photoionized nebulae --both {\hii} regions and planetary nebulae (PNe)-- optical recombination lines (ORLs) provide abundance values that are systematically larger than those obtained using collisionally excited lines (CELs) \citep[see e.~g.][, and references therein]{peimbertpeimbert06, liu12}. This is known as the {\it abundance discrepancy problem}. It is one of the major unresolved problems in nebular astrophysics, being known of for more than seventy years \citep{wyse42}, and has far-reaching consequences on the measurement of abundances throughout the Universe, most often measured using CELs from ionized gas.

The abundance discrepancy factor ({\adf}) is defined as the ratio between the abundances derived from ORLs and CELs and is usually between 1.5 and 3 \citep[see e.~g.][]{garciarojasesteban07,liu12}, but in PNe it has a significant tail extending to much larger values. It has recently been shown that the largest abundance discrepancy factors (up to $\sim$300), are reached in PNe with close binary central stars that have undergone a common envelope (CE) phase \citep{corradietal15, jonesetal16}. The data presented in these papers support the hypothesis that two different gas phases may coexist \citep{liuetal06}: hot, standard metallicity gas at $\sim$10000 K where the CELs can be efficiently excited, and much cooler ($\sim$1000 K) gas clumps with a highly enhanced content of heavy elements (which is the cause of the cooling) where only ORLs form.  As this dual nature is not predicted by mass loss theories, these results have added a new, unexpected ingredient to  the abundance discrepancy problem: very high discrepancy factors should be explained in the framework of (interacting) binary evolution. Some of the proposed explanations are naturally linked to binarity, such as for instance high-metallicity nova-like ejecta \citep{wessonetal08}. Other appealing explanations involve the presence of planetary debris that survived the whole evolution of the central star, or the tidal destruction, accretion and ejection of Jupiter-like planets \citep[see the discussion in][]{corradietal15}, but it is difficult at this stage to favor any scenario because observational constraints are still very scarce. 

The existing long-slit  studies point out that the plasma emitting in {\oii} ORLs (the brightest metal ORLs in the optical spectrum) is generally more concentrated in the central parts of the nebulae than the CEL emitting zones of the same ion \citep[e.~g.][]{liu12,corradietal15}. However, no direct imaging of the gas component where the bulk of the metal ORLs is emitted has ever been attempted, mainly because {\oii} ORLs are faint and suitable filters are generally not available.


NGC~6778 ($\alpha$ = 19$^h$ 18$^m$ 24.94$^s$, 
$\delta$ = $−$01$\arcdeg$ 35$\arcmin$ 47.641$\arcsec$, PN\,G034.5−06.7) 
was found to show unusually strong ORLs reflecting a high {\adf} value \citep[average value $\sim$20, see][]{jonesetal16}. It hosts a close binary central star with an orbital period of 3.7 hours, making it one of the shortest period binary central stars known \citep{miszalskietal11}. 

In the present work, we take advantage of the large collecting area of the 10.4m Gran Telescopio Canarias (GTC) telescope and its Blue Tunable Filter (BTF) to study the 2D emission profiles of the {\oii} ORL and {\foiii} CELs for the first time. 

\section{Observations and data reduction}\label{sec:obs}

\subsection{OSIRIS-GTC data}

We obtained narrow band imagery of NGC\,6778 using the Optical System for Imaging and Low Resolution Integrated Spectroscopy (OSIRIS, \citealp{cepaetal00, cepaetal03}) on the 10.4m Gran Telescopio Canarias (GTC), at the Observatorio del Roque de los Muchachos on La Palma. The OSIRIS detector consists of a mosaic of two Marconi CCDs, each with 2048 $\times$ 4096 pixels and a total  unvignetted field of view of $7.8 \times 7.8$ arcmin$^2$, giving a plate scale of 0.127 arcsec/pix. We selected the 2$\times$2 binning mode and the 200 kHz readout mode. 

We used the OSIRIS Blue Tunable Filter (BTF) for our observations. It consists of a Fabry-Perot etalon which allows narrow band imaging at a selected wavelength. However, Fabry-Perot etalons cause a radial wavelength dependence in the form of ``rings'' of constant wavelength centred on the optical axis. Therefore, in order to sample the PN {\oii} emission  at 4649+50 \AA, the BTF tuning was displaced to 465.4$\pm$0.2 nm, with an effective bandpass of 0.85 nm. This ensures that the wavelength selection is the desired one to isolate {\oii} emission when the target is located  1 arcmin from the TF optical center. There was no need to scan the TF as the nebula is sufficiently small ($<$25 arcsec diameter excluding the faintest collimated outflows) that the wavelength variation from one side to the other of the nebula is $<$3 \AA\ at the selected location.

Data for NGC\,6778 were taken in service mode during the night of 2015 August 15 under Director's Discretionary Time. A total integration time of 3 $\times$ 600 s was obtained, with small offsets to remove the sky background contribution. The OSIRIS BTF images were bias- and flat-field corrected in the usual way using IRAF\footnote{IRAF is distributed by the National Optical Astronomy Observatory, which is operated by the Association of Universities for Research in Astronomy (AURA) under a cooperative agreement with the National Science Foundation.}. Sky emission was removed by producing an image containing only the sky emission from the series of dithered images. This sky image was then subtracted from the original images of the source, resulting in images cleaned from any background and sky emission. To save telescope time, owing to the large exposure times needed to obtain the desired S/N ratio in the {\oii} lines, no attempt was made to subtract the negligible nebular continuum. 

Additionally, we show {\foiii} $\lambda$5007, {\ha} and {\fnii} $\lambda$6548 narrow-band images taken from \citet{guerreromiranda12} (see Sect.~\ref{results}). 
All the images were registered to the {\foiii} image pixel scale (0.189 arcsec pix$^{-1}$) and astrometrized using the 2MASS catalog and the script  designed by \citet{sanabria15}.

\subsection{VIMOS-VLT data}

\begin{figure}[!ht]
\includegraphics[angle=0,scale=0.4]{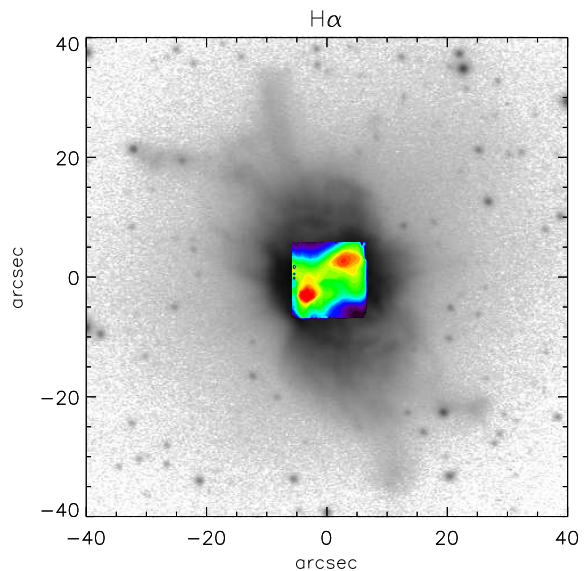}
\caption{The VIMOS {\ha} map (color image) overlaid on ALFOSC-NOT narrow band image showing the area covered by the VIMOS IFU (13.2$\times$13.2 arcsec$^2$). The ALFOSC-NOT image was taken from \citet{guerreromiranda12}. }\label{fig1}
\end{figure}

NGC\,6778 was observed with VIMOS \citep{lefevreetal03}, on the ESO-VLT UT3 Melipal on the night of 2007 September 14. VIMOS is equipped with an integral field unit containing 6400 fibers
and has a changeable scale on the sky that was set to 0.33 arcsec per fiber. This provided a sky coverage of 13.2$\times$13.2 arcsec$^2$ with an image formed by an array of 40$\times$40 fibers. We obtained observations in the VIMOS high-resolution mode, providing a reciprocal dispersion of  0.6 \AA\ pix$^{-1}$ and a usable range from 3900 to 7400 \AA\ adding the blue and red grisms of the spectrograph. The object was observed with an exposure
time of 300 s in both the red and blue configurations. The reduction was performed
with the VIMOS pipelines available at the instrument website\footnote{http://www.eso.org/sci/facilities/paranal/instruments/vimos/}. For sky subtraction we observed fields offset by 40 arcsec. The sky level was estimated as an average for each pixel along the dispersion axis and then subtracted from the object data cube. We also applied a correction for the differential atmospheric extinction using the central star of NGC\,6778 in the data cube and performing the necessary shifts along the dispersion axis.
To extract the final VIMOS emission line maps we used a set of IDL routines which fit Gaussian profiles to the lines of interest in each spaxel in the data cube. In Fig.~\ref{fig1} we show the area covered by the VIMOS IFU by overlaying the VIMOS map taken in the light of {\ha} on an ALFOSC-NOT narrow-band image centered on the same line from \citet{guerreromiranda12}. The IFU was positioned to cover the central 13.2$\times$13.2 arcsec$^2$ of the nebula. 

\section{Results and discussion}\label{results}

\subsection{OSIRIS-GTC and ALFOSC-NOT images}\label{images}

\begin{figure*}[!ht]
\centering
\includegraphics[angle=0,scale=0.3]{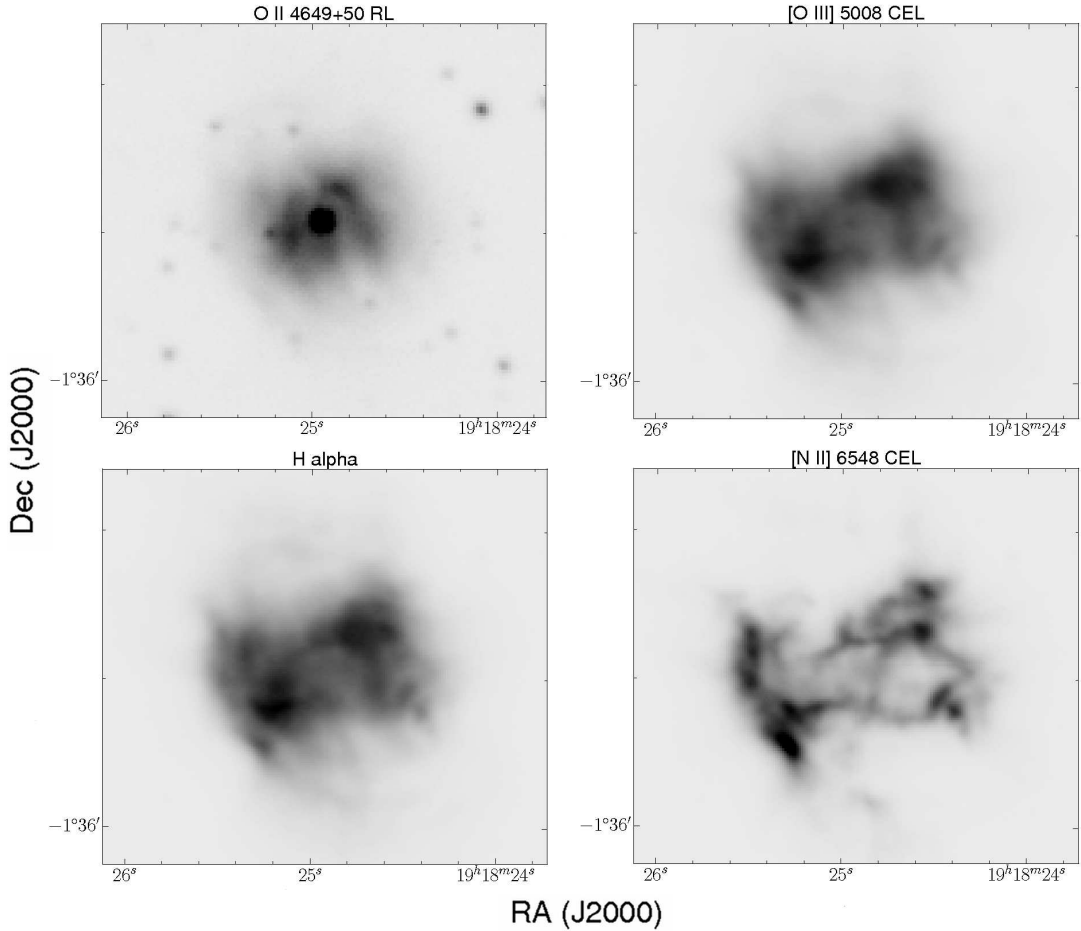}
\caption{OSIRIS-GTC Tunable-filter image of NGC\,6778 in the {\oii} 4649+51 ORLs (upper left), compared to the ALFOSC-NOT images from \citet{guerreromiranda12} in the {\foiii} 5007 CEL (upper right), {\ha} (lower left) and {\fnii} 6548 (lower right).
All the images were registered to the {\foiii} image pixel scale (0.189 arcsec/pixel) and astrometrized using the 2MASS catalog. It is worth mentioning that {\oii} and {\foiii} emission comes from the same ion: O$^{2+}$.}\label{fig2}
\end{figure*}

\begin{figure}[!ht]
\includegraphics[angle=0,scale=0.45]{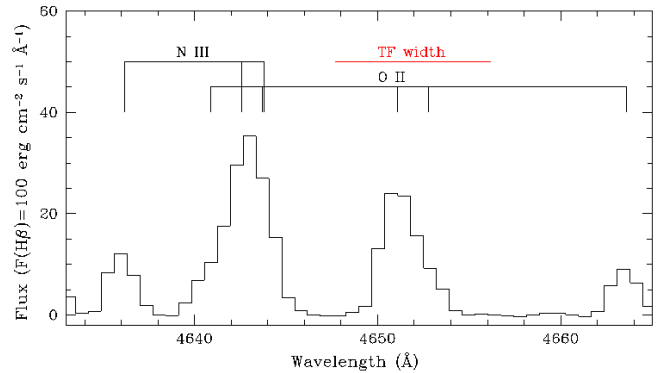}
\caption{Section of the VLT-FORS2 observed spectrum (black line) of NGC\,6778 presented in \citet{jonesetal16} where the {\oii} ORLs, as well as some {\niii} lines, have been highlighted. The TF width is shown as a red line.}\label{fig3}
\end{figure}

In the upper-left panel of Fig.~\ref{fig2} we show the OSIRIS-GTC BTF image centered on the {\oii} $\lambda\lambda$4649+51 ORLs showing the spatial distribution of O$^{2+}$ from these lines. In the upper-right panel, we show the narrow-band {\foiii} $\lambda$5007 CEL image from \citet{guerreromiranda12}. 
The lower panels present narrow-band images in {\ha} (left) and {\fnii} $\lambda$6548 CEL (right) from the same authors. It is known that in several PNe the {\oii} ORLs are contaminated by fluorescent {\niii} lines and/or {\ciii} ORLs. In their deep VLT-FORS2 spectrum of NGC\,6778, \citet{jonesetal16} did not report any contamination in the wavelength range covered by our tunable filter. In Fig.~\ref{fig3}, we show a section of the observed spectrum of NGC\,6778 around the multiplet 1 of {\oii}. It is clear that there is no emission from the {\ciii} $\lambda$4647.42 line, which is the brightest one of the multiplet 1 of {\ciii}. Therefore, all the emission covered by the tunable filter comes purely from {\oii} recombination lines. 

From inspection of Fig.~\ref{fig2}, it is clear that the {\foiii} $\lambda$5007 and {\ha} emissions share a very similar spatial distribution in the central parts of NGC\,6778 \citep[see also the discussion in ][]{guerreromiranda12}. On the other hand, the {\fnii} $\lambda$6548 emission has a very different spatial distribution. \citet{guerreromiranda12} showed that the innermost region of NGC\,6778 can be described as a distorted and fragmented {\fnii}-bright equatorial ring aligned along a line close to the east-west direction. This {\fnii} CEL innermost region is characterized by the presence of filamentary low-ionization structures (LIS). Their origin, however, remains unknown but has been linked to post-CE nebulae through photoionization of neutral material deposited in the orbital plane during the CE phase \citep[see ][and references therein]{miszalskietal11, manicketal15}. 

The most important finding from inspection of Fig.~\ref{fig2} is that in NGC\,6778 the location of the O$^{2+}$ ions producing the {\oii} ORLs at 4649+50 \AA\ does not match the spatial distribution of the same ions emitting in the {\foiii} $\lambda$5007 CEL. The {\oii} ORL emission is concentrated inside the {\foiii} CEL emission or {\hi} emission. This is the first time that the different distribution of {\oii} ORL and nebular {\foiii} CEL emission has been shown using direct imaging. This finding clearly supports the hypothesis of the existence of two separate plasmas, with the additional indication that they are not well mixed, perhaps because they were produced in distinct ejection events. Previous works based on long-slit and 2D spectrophotometry in partial regions of PNe found similar results. Hence, this behaviour seems to be a standard characteristic of the ORL emission in PNe with high {\adfs} \citep[see e.~g.][]{liuetal00, tsamisetal08, corradietal15}. 

\subsection{VIMOS monochromatic emission line maps}

\begin{figure}[!ht]
\includegraphics[angle=0,scale=0.45]{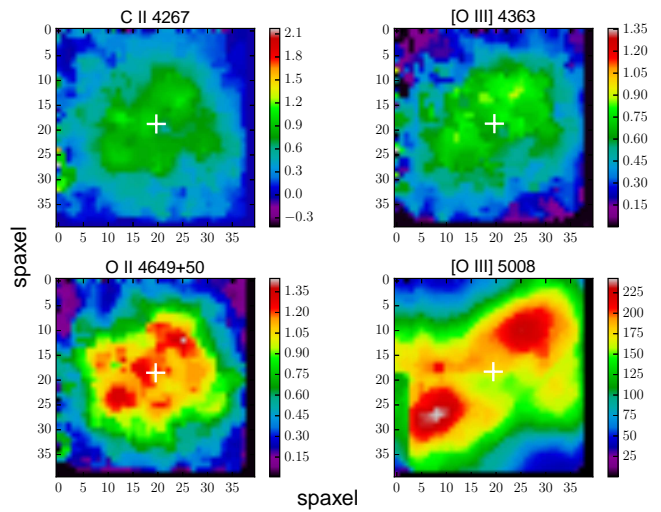}
\caption{VIMOS-VLT emission line maps in several emission lines. Left column: {\cii} $\lambda$4267 and {\oii} $\lambda\lambda$4649+50 ORLs; 
right column:  auroral {\foiii} $\lambda$4363 and nebular $\lambda$5007 CELs.
The maps area is 13.2$\times$13.2 arcsec$^2$ and the orientation is north up and east left. Position of the central star is marked with a white cross.}\label{fig4}
\end{figure}

We have analyzed the morphology of the emission coming from various other lines on monochromatic emission line maps taken with VIMOS-VLT. In the left column of Fig.~\ref{fig4}, we show the {\cii} $\lambda$4267 and {\oii} $\lambda\lambda$4649+50 maps which characterize the spatial distribution of pure C$^{2+}$ and O$^{2+}$ recombination emission. Although the {\cii} emission has a rather low contrast relative to the background, its global morphology resembles the {\oii} $\lambda\lambda$4649+50 emission, showing local peaks at very similar spaxels. The right column of Fig.~\ref{fig4} shows the resemblance of the auroral {\foiii} $\lambda$4363 CEL and the nebular {\foiii} $\lambda$5007 CEL emission maps. The morphologies of {\oii} $\lambda$4649+50 and {\foiii} $\lambda$5007 emissions are remarkably different, and are consistent with what is found from the GTC and NOT imagery (see Fig.~\ref{fig1}). 

On the other hand, it is clear that the {\foiii} $\lambda$4363 and {\foiii} $\lambda$5007 CEL emissions have a very different spatial distribution, but  {\foiii} $\lambda$4363 resembles the {\oii} ORL emission.  This causes a radially increasing {\foiii}5007/4363 flux ratio, which may be interpreted as being due to a markedly negative electron temperature gradient \citep[][estimated a negative gradient of more than 1000 K, see their Fig.~5]{jonesetal16}. This effect has been found in other PNe \citep[see][for the PNe NGC\,6153 and NGC\,6720, respectively]{liuetal00, garnettdinerstein01}. However, the presence of extremely high-density clumps ({\elecd}$>$10$^5$ cm$^{-3}$) in the inner regions would cause the quenching of nebular {\foiii} lines in the central part of the PN owing to collisional de-excitation effects. The effect of high-density ionized gas in the determination of chemical abundances from ORLs and CELs has been studied in detail in i.~e. the Orion nebula, by \citet{tsamisetal11} and \citet{mesadelgadoetal12}.
This would affect the determination of physical conditions ({\elecd} and {\te}) and abundances from CELs, creating a spurious temperature gradient.  Moreover, if these clumps are H-poor and only contribute a few percent of the total H mass of the PN, as it has been proposed for PN, they would not be seen in {\hi} images. Therefore, our observations support  the presence of a metal rich, dense component in the inner regions of NGC 6778, where ORL emission mainly form, in addition to the "standard", low density and H-rich gas component where CELs are emitted. Other than its location in the inner nebular regions, which may indicate a younger origin, the morphology of the ORL emitting gas does not provide any other clues to its origin.
Detailed 3D photoionization models are therefore needed to fully constrain the nebula and shed further light on its formation.

All these results are in very good agreement with what was found by \citet{jonesetal16},  although they were limited to a 1-D portion of the nebulae as selected in their long-slit FORS2 observations.

Finally, some kinematical studies point to the presence of a colder plasma where ORLs are preferentially emitted. From very high spectral resolution observations (R=150,000) of two moderate to high {\adf} PNe (NGC\,7009 with {\adf}$\sim$5 and  NGC\,6153 with {\adf}$\sim$10) \citet{barlowetal06} found that the full width at half-maximum (FWHM) of {\oii} ORLs were significantly smaller than that shown by {\foiii} CELs. They also found that auroral and nebular {\foiii} lines in NGC\,6153 showed different velocity profiles, which implies large temperature fluctuations. Such large temperature fluctuations can also contribute to the observed {\adf} \citep[see][and references therein]{garciarojasesteban07}. \citet{richeretal13} made a detailed kinematical study of ORLs and CELs in a limited region of PN NGC\,7009. They found that ORLs exhibit different kinematics to CELs but no clear conclusion could be drawn because of the limited spatial coverage. \citet{otsukaetal10} analyzed high resolution (R$\sim$33,000) spectra of the halo PN BoBn~1. They found a relatively low {\adf} ($\sim$2.5) and  {\oii} and {\neii} ORLs expansion velocities lower than that of {\foiii} and {\fneiii} CELs, which they attributed to different thermal motions. It is worth mentioning that in the case of NGC\,6778, where the {\oii} ORL emission comes from the innermost zones of the PN, the kinematics can be affected by the expansion velocity field combined with the ionization stratification of the gas in the PN. This would result in narrower {\oii} ORLs compared to nebular {\foiii} CELs.

\section{Conclusions}

We present direct imaging in the light of the {\oii} $\lambda\lambda$4649+50 ORLs for the first time in a PN. We have found that the distribution of the O$^{2+}$ emission coming from these lines in NGC\,6778 is more centrally concentrated than the emission coming from either the strongest {\foiii} $\lambda$5007 CEL or the {\ha} RL. Also, from 2D reconstructed VIMOS monochromatic emission line maps, we find that the auroral {\foiii} $\lambda$4363 CEL emission resembles the {\oii} ORLs emission but not the nebular {\foiii} emission. This may be due to presence of a high-density, H-poor gas component in the inner regions of the nebula, which could be responsible for the bulk of the {\oii} emission and cause the abundance discrepancy.  
However, we do not have enough information to generalize the scenario of metal-rich inclusions to explain the bulk of low-to-moderate ADF PNe or {\hii} regions. Moreover, the effect of high-velocity flows, such as in HH objects, and/or high-density inclusions, such as proplyds, has definitely something to say in the abundance discrepancy problem in {\hii} regions.

The physical origin of the enhancement of the {\oii} emission in the central parts of this PN is still elusive, but should almost certainly be linked to the binary evolution of its central star. 
Additional observations are needed to put this important result on solid ground and further progress in our  understanding of the origin of the phenomenon. Determining the spatial distribution of the ORLs and CELs, and confronting their dynamics in several PNe with high {\adf} values would help constrain the geometry and ejection mechanism of the metal rich (H-poor) component.

\acknowledgments

This work is based on observations obtained with the 10.4m GTC, operating on the island of La Palma at the Spanish Observatories of the Roque de Los Muchachos of the Instituto de Astrof\'\i sica de Canarias.  We thank the anonymous referee for his/her comments. This research has been supported by the Spanish Ministry of Economy and Competitiveness (MINECO) under grants AYA2012-35330, AYA2011-22614, and AYA2012-38700. JGR acknowledges support from Severo Ochoa excellence program (SEV-2011-0187) postdoctoral fellowship. PRG was supported by a Ram\'on y Cajal fellowship (RYC2010-05762). JGR acknowledges fruitful discussions with C. Esteban.

{\it Facilities:} \facility{GTC (OSIRIS)} \facility{VLT (VIMOS)}.











\begin{thebibliography}{}

\bibitem[Barlow et al.(2006)]{barlowetal06}
Barlow, M.~J., Hales, A.~S., Storey, P.~J., et al. 2006, in IAUSymp. 234, PlanetaryNebulae in our Galaxy and Beyond,
eds. M.~J. Barlow \& R.~H. M\'endez (Cambridge: Cambridge Univ. Press), 367

\bibitem[\protect\citeauthoryear{Cepa et al.}{2000}]{cepaetal00} 
Cepa J., et al., 2000, SPIE, 4008, 623 

\bibitem[\protect\citeauthoryear{Cepa et al.}{2003}]{cepaetal03} 
Cepa J., et al., 2003, SPIE, 4841, 1739 

\bibitem[Corradi et al.(2015)]{corradietal15}
Corradi, R. L. M., Garc\'{\i}a--Rojas, J.,  Jones, D., \& Rodr{\'{\i}}guez--Gil, P. 2015, \apj, 803, 99


\bibitem[Garc\'\i a--Rojas \& Esteban(2007)]{garciarojasesteban07}
Garc\'\i a--Rojas, J., \& Esteban, C., 2007, \apj, 670, 457

\bibitem[Garnett \& Dinerstein(2001)]{garnettdinerstein01}
Garnett, D.~R., \& Dinerstein, H.~L., 2001, \apj, 558, 145

\bibitem[Guerrero \& Miranda(2012)]{guerreromiranda12}
Guerrero, M.~A., \& Miranda, L.~F., 2012, A\&A, 539, A47

\bibitem[Jones et al.(2016)]{jonesetal16}
Jones, D., Wesson, R., Garc\'{\i}a-Rojas, J., Corradi, R. L. M., \& Boffin, H.~M.~J. 2016, \mnras, 455, 3263

\bibitem[Le F\`evre et al.(2015)]{lefevreetal03}
Le F\`evre, O., Saisse, M., Mancini, D., et al. 2003, in SPIE Conf. Ser. 4841, eds. M. Iye ,\& A. F. M. Moorwood, 1670

\bibitem[Liu(2012)]{liu12} 
Liu, X.-W. 2012, in IAU Symp. \#283, Planetary Nebulae: An Eye to the Future, 
Manchado, A. Stanghellini, L., \& Sch\"onberner, D. eds., p. 131

\bibitem[Liu et al.(2000)]{liuetal00}
Liu, X.~-W., Storey, P.~J., Barlow, M.~J., Danziger, I.~J., Cohen, M., \& Bryce M. 2000, \mnras, 312, 585

\bibitem[Liu et al.(2006)]{liuetal06}
Liu, X.~-W., Barlow, M.~J., Zhang, Y., Bastin, R.~J., \& Storey P.~J. 2006, \mnras, 368, 1959

\bibitem[Manick et al.(2015)]{manicketal15}
Miszalski, B., Jones, D., Rodr{\'{\i}}guez--Gil, P., Boffin, H.~M.~J., Corradi, R. L. M., \& Santander--Garc\'{\i}a, M. 2011, A\&A, 531, A158

\bibitem[Mesa-Delgado et al.(2012)]{mesadelgadoetal12}
Mesa-Delgado, A., N\'u\~nez-D\'{\i}az, M., Esteban, C. et al. 2012, \mnras, 426, 614

\bibitem[Miszalski et al.(2011)]{miszalskietal11}
Manick, R., Miszalski, B.,\& McBride, V. 2015, \mnras, 448, 1789


\bibitem[Otsuka et al.(2010)]{otsukaetal10}
Otsuka, M., Tajitsu, A., Hyung, S., \& Izumiura, H. 2010, \apj, 723, 658

\bibitem[Peimbert \& Peimbert(2006)]{peimbertpeimbert06}
Peimbert, M., \& Peimbert, A. 2006, in IAU Symp. \#234. Barlow, M.~J., \& M\'endez, R.~H.  eds., p. 227

\bibitem[Richer et al.(2013)]{richeretal13}
Richer, M.~G., Georgiev, L., Arrieta, A., \& Torres-Peimbert, S. 2013, \apj, 773, 133

\bibitem[Sanabria(2015)]{sanabria15} 
Sanabria, J.~J. 2015, Master Thesis, Valencian International University

\bibitem[Tsamis et al.(2008)]{tsamisetal08}
Tsamis, Y.~G., Walsh, J.~R., P\'equignot, D.,  et al. 2008, \mnras, 386, 22

\bibitem[Tsamis et al.(2011)]{tsamisetal11}
Tsamis, Y.~G., Walsh, J.~R., V\'{\i}lchez, J.~M., \& P\'equignot, D. 2011, \mnras, 412, 1367

\bibitem[Wesson et al.(2008)]{wessonetal08}
Wesson, R.,  Barlow, M.~J., Liu, X.-W., et al. 2008, \mnras, 383, 1639

\bibitem[Wyse(1942)]{wyse42}
Wyse, A.~B. 1942, \apj, 95, 356

\end{thebibliography}
\end{document}